\documentclass[12pt,preprint]{aastex}

\def\bS{{\bf J}}
\def\bL{{\bf L}}
\def\<{\langle}
\def\>{\rangle}
\def\msun{{\rm\,M_\odot}}

\def\au{{\rm\,AU}}

\def\eps{\epsilon}

\def\sE{{\mathcal{E}}}
\def\order{{\mathcal{O}}}

\def\gta{\gtrsim}
\shortauthors{Gammie et al.}
\shorttitle{Spin Evolution}
\begin{document}
\title{Black Hole Spin Evolution}

\author{Charles F. Gammie\altaffilmark{1,2}, Stuart L.
Shapiro\altaffilmark{1,2,3}, and Jonathan C. McKinney\altaffilmark{2}}

\affil{Department of Physics, University of Illinois at Urbana-Champaign\\
1110 West Green St., Urbana, IL 61801, USA; }

\email{gammie@uiuc.edu, shapiro@astro.physics.uiuc.edu, jcmcknny@uiuc.edu}

\altaffiltext{1}{Department of Astronomy, University of Illinois at
Urbana-Champaign, Urbana, IL 61801}
\altaffiltext{2}{Center for Theoretical Astrophysics, University of
Illinois at Urbana-Champaign, Urbana, IL 61801} 
\altaffiltext{3}{National Center for Supercomputing Applications,
University of Illinois at Urbana-Champaign, Urbana, IL 61801} 

\begin{abstract}

We consider a subset of the physical processes that determine the spin
$j \equiv a/M$ of astrophysical black holes.  These include: (1) Initial
conditions.  Recent models suggest that the collapse of supermassive
stars are likely to produce black holes with $j \sim 0.7.$ (2) Major
mergers.  The outcome of a nearly equal mass black hole-black hole
merger is not yet known, but we review the current best guesses and
analytic bounds. (3) Minor mergers.  We recover the result of Blandford
\& Hughes that accretion of small companions with isotropically
distributed orbital angular momenta results in spindown, with $j \sim
M^{-7/3}$.  (4) Accretion.  We present new results from fully
relativistic magnetohydrodynamic accretion simulations.  These show
that, at least for one sequence of flow models, spin equilibrium ($dj/dt
= 0$) is reached for $j \sim 0.9$, far less than the canonical value
$0.998$ of Thorne that was derived in the absence of MHD effects. This
equilibrium value may not apply to all accretion flows, particularly
thin disks.  Nevertheless, it opens the possibility that black holes
that have grown primarily through accretion are not maximally rotating.

\end{abstract}


\section{Introduction}

The massive, dark objects observed in the centers of galaxies (e.g.
\citealt{miy95,mag98}) and the stellar-mass compact objects observed in
binary systems systems \citep{mr03} are most readily interpreted as
black holes.  Alternative models require the introduction of exotic
physics \citep{bahc90} or modification of Einstein's equations for the
gravitational field (recently \citealt{dp03}).  More conventional models
such as clusters of compact stars are strongly constrained by
observations.  Most remarkably, proper motion and radial velocity
studies of stars near the putative black hole in Sgr A$^*$
\citep{sch02,gez02} require that approximately $3 \times 10^6 \msun$ be
concentrated within a region $120 \au$ in radius.  There is no stable
configuration of normal matter with such a large mass in such a small
volume; cluster lifetimes are too short \citep{mao98}.  Black holes are
therefore the ``most conservative'' model for massive dark objects and
galactic black hole candidates (hereafter GBHCs).

Black hole solutions of Einstein's equations have three parameters: mass
$M$, spin $\bS$, and charge $Q$ (by the ``no-hair,'' or uniqueness,
theorem; see \citealt{wal84}).  Of these, $Q$ is likely to be negligible
in astrophysical contexts because electric charge is shorted out by the
surrounding plasma \citep{bz77}.  Thus while much of the variation in
the observational appearance of black holes is likely due to variation
in external parameters such as the angle between black hole spin vector
and line of sight, the gas accretion flow geometry and accretion rate
$\dot{M}$, and other environmental factors, some might also be due to
variation in black hole spin $j \equiv J/M^2 = a/M$.

Several features of supermassive black holes (hereafter SMBHs) and GBHCs
have been interpreted as evidence for black hole spin:

(1) Some SMBHs and GBHCs show broad, skewed Fe K$\alpha$ lines, for
example in MCG-6-30-15 \citep{tan95,fab02}, Cyg X-1 \citep{mil02a}, and
XTE J1650-500 (\citealt{mil02b}; for a review see \citealt{rn02}).  If
one assumes that these lines originate in plasma on nearly circular,
equatorial geodesics within a few $M$ of the black hole, then the line
shape is sensitive to the spin of the hole (e.g. \citealt{lao91}).
Within the context of this model rotating holes are required to explain
the observed red wing of the line.

(2) The ratio $R$ of observed quasar radiative energy per unit comoving
volume to the current mass density of black holes is directly related to
the mean radiative efficiency $\eps$ of accretion onto black holes
\citep{sol82}: $\eps > R$.  Estimates suggest $R > 0.1$
\citep{yt02,erz02}.  If one assumes that accretion occurred through a
classical thin disk in which the binding energy of the innermost stable
circular orbit (hereafter ISCO) determines $\eps$ \citep{bar70}, then $R
> 0.1$ requires $j > 0.67$.

(3) In GBHCs, quasi-periodic oscillations (QPOs) are observed in X-ray
light curves at frequencies ranging from a fraction of a Hz to $450$ Hz
in GRO J1655-40 \citep{str01}.  Assuming that these QPO frequencies are
bounded above by the rotation frequency of the ISCO and that the QPO is
not an overtone, one can place a limit on the mass and spin of the black
hole.  In GRO J1655-40, $95\%$ confidence limits on the mass
\citep{sha99} require $M > 5.5 \msun$, or $j  > 0.15$ \citep{str01}.  A
physical or phenomenological model for the QPO can provide more
stringent constraints, but requires additional assumptions.

(4) The shape of the X-ray continuum from an accreting black hole may
depend on the spin.  Calculating an expected continuum requires the
black hole mass, spin, flow geometry (usually, but not always, a thin
disk) and a model for the accretion flow atmosphere.  Models have been
applied to a number of objects by, e.g., \cite{zcc97} and \cite{gme01},
and usually suggest $j \sim 1$.

%
%
%

This list is necessarily incomplete, and in each case the evidence for
black hole spin is open to debate.  Models for the dynamics of the
plasma surrounding the black hole and its radiative properties must be
invoked.  These models describe an intrinsically complex physical system
and use approximations of unknown accuracy.  Future calculations,
particularly numerical models of the accretion flows, may help reduce
the uncertainties.  Analysis of gravitational waveforms emitted by
perturbed black holes undergoing mergers can reveal their masses and
spins and may prove less ambiguous once these signals can be measured
reliably \citep{tho95,flan98,drey03}.

Given the existing evidence for black hole spin, it is useful to
consider the physical processes governing spin evolution.  In this paper
we consider initial conditions (\S\ref{initial}), mergers with black
holes of comparable mass (\S\ref{major}), mergers with smaller objects
(\S\ref{minor}), and accretion (\S\ref{accretion}), then summarize our
results in \S\ref{finale}.  Throughout the paper we adopt geometrized
units and set $G = c = 1$.

\section{Initial Conditions}\label{initial}

Nonprimordial black holes form from gravitational collapse and in
general are born with nonzero spin.  If subsequent accretion is
negligible, the initial spin state will be preserved and the black hole
spin will be determined by the dynamics of the initial collapse.

If the initial collapse occurs from a massive star, the spin depends on
the angular momentum profile of the progenitor star and the
(magneto)hydrodynamics of core collapse in evolved, spinning stars. The
dependence is not fully understood at this time, although detailed
Newtonian simulations of the collapse of spinning stars with $M \lesssim
300 M_{\odot}$  have been performed and suggest how spinning black holes
may arise during core collapse (see, e.g. \citealt{heg02} for a review
and references). The simulations show that the fate of the collapse
depends critically on the mass, spin, metallicity and magnetic field of
the progenitor, as well as details of the equation of state and neutrino
transport, so it is not surprising that the issue is not resolved.

There are also results for idealized versions of the general
relativistic collapse problem.  \cite{shib02} followed the collapse of a
marginally unstable supermassive star in full general relativity.  They
considered the case of a uniformly rotating star supported by radiation
pressure and spinning at the mass-shedding (maximal spin) limit.
(Mass-shedding is the likely situation by the time the star has cooled
and contracted quasistatically to the point of onset of collapse
\citep{baum99}, provided that the star can sustain solid body rotation
during the contraction phase.)  They found that the final object is a
Kerr--like black hole surrounded by a disk of orbiting gaseous debris.
The final black hole mass and spin were determined to be  $M_{\rm h}/M
\approx 0.9$ and $J_{\rm h}/M^2_{\rm h} = j = a/M \approx 0.75$, for an
arbitrary progenitor star mass $M$.  The remaining mass goes into the
disk of mass $M_{\rm disk}/M \approx 0.1$.  In fact, the final black
hole and disk parameters can be calculated analytically from the initial
stellar density and angular momentum distribution \citep{shap02}.  The
results obtained here apply to the collapse of {\it any} marginally
unstable $n \approx 3$ polytrope spinning uniformly at mass-shedding.
Hence these results may be applicable to core collapse in very massive
stars $\gtrsim 300 {\rm M_{\odot}}$ and to collapsar models of
long-duration gamma-ray bursts \citep{mw99}.  This work suggests,
therefore, that black holes formed during core collapse of massive stars
should be born rapidly rotating, but well below the Kerr limit.

Black holes may also arise in other dynamical scenarios, like the
coalescence of binary neutron stars. The most detailed calculations of
binary neutron star mergers in full general relativity are the
hydrodynamic simulations of \cite{su02}. They considered the coalescence
of irrotational binaries (physically the most likely case),  modeled as
equal-mass polytropes with adiabatic indices $\Gamma = 2$ and $2.25$,
and considered a range of initial masses below the maximum mass limit.
They followed the merger from the ISCO to coalescence. For intermediate
mass stars, the merged remnant is a differentially rotating,
``hypermassive'' neutron star. For high mass stars, the remnant is a
rotating black hole. For the binaries which formed black holes, the
precollapse $J/M^2$ ranged from $0.9$ -- $1.0$, with the higher values
associated with the smaller masses. The black hole products have $J/M^2
\sim 0.8$ -- $0.9$ due to the loss of $\sim 10\%$ of the initial angular
momentum through gravitational radiation.  In these calculations most of
the mass is conserved and goes into the black hole, and no disk forms
about the hole.

\section{Spin-up By Major Mergers}\label{major}

Black hole spin will also change when a black hole merges with a black
hole of comparable mass (a ``major merger'').  The outcome of this
merger and the partitioning of energy and angular momentum between
internal and radiative degrees of freedom is an area of active research.
Here we summarize some current estimates.

Consider binary black holes inspiralling due to gravitational radiation
from initially circular orbits (gravitational radiation reduces the
eccentricity of the orbits on a timescale short compared to the orbital
evolution timescale).  Once the black holes reach the
ISCO, they will plunge together and merge on an orbital
timescale. The location of the ISCO, as well as the global parameters
characterizing the binary at this critical separation, are not known to
high precision except for black holes with test-particle companions.
Several different approaches have been formulated and have yielded
approximate solutions (for a recent review and references, see \citealt{bs03}).
These range from high-order post-Newtonian calculations to fully
nonlinear numerical solutions of the initial value vacuum Einstein
equations. We compare some of the available results for nonspinning
black holes in Table 1, adapted from \cite{bs03}.
                                                                                
In constructing this table, we identify the mass of each black hole with
the irreducible mass
\begin{equation} 
M_{\rm BH} = M_{\rm irr} = \left(\frac{A}{16 \pi}\right)^{1/2},
\end{equation}
where $A$ is the proper area of the black hole's event horizon
\citep{chr70}.  The binding energy can then be defined as
\begin{equation}
E_b = M - 2 M_{\rm BH},
\end{equation}
where $M$ is the total (ADM) mass of the system measured at large
distance from the holes (see, e.g., \citealt{my74}).  Values for the
nondimensional binding energy $\bar E_b \equiv E_b/\mu$, the orbital
angular velocity $\bar \Omega \equiv m \Omega$ and the angular 
momentum $\bar J \equiv J/(\mu m)$ at the ISCO are listed.  Here $\mu$
is the reduced mass, $\mu = M_{\rm BH}/2$, and $m$ is the sum of the black
hole masses, $m = 2 M_{\rm BH}$.  The fractional losses from the system of
mass and angular momentum due to gravitational radiation emission during
the plunge are expected to be small (see, e.g., \citealt{kh99}), but precise
values await more reliable relativistic calculations. Meanwhile, a reasonable
first approximation is to assume that the final black hole
will have a mass and angular momentum nearly equal to the binary system
at the ISCO. With this assumption, the spin parameter of the final black
hole is given by
\begin{equation}
\label{spin}
{J \over M^2} = {\bar J \over {4 \left [1 + {\bar E_b \over 4} \right ]^2}}
\end{equation}
Results of the numerical calculations of \cite{coo94}, \cite{bau00}, and
\cite{ggb02} are tabulated, as well as the third-order Post-Newtonian
results of \cite{djs00}.  The final black hole spin parameter computed
according to equation (\ref{spin}) is listed in the fifth column in the
table. A strict upper limit to the final spin parameter, $(J/M^2)_{\rm
max}$, is provided by the black hole area theorem and is listed in the
sixth column for comparison (see equation (\ref{max}) and discussion
below.) In the table we also include the analytical values for a test
particle orbiting a Schwarzschild black hole, $\bar E_b = \sqrt{8/9} - 1
= -0.0572$, $\bar J = 2 \sqrt{3} = 3.464$, $\bar \Omega = 1/6^{3/2} =
0.0680$, with $J/M^2$ evaluated according to equation (\ref{spin}) and
$(J/M^2)_{\rm max}$ calculated according to equation (\ref{max}).

Despite the differences in the ISCO calculations (for a recent detailed
comparison, see \citealt{cook03}), the values obtained for the expected
final spin parameter based on mass and angular momentum conservation are
all comparable and high, $J/M^2 \gtrsim 0.8$.  Note that the range of
these estimated final spin parameters is far narrower than the range of
calculated ISCO orbital frequencies.  Note also that these results are
for nonspinning holes.

For spinning black holes one knows from test particle orbits that the
location of the ISCO depends strongly on spin.  Most numerical
treatments of this problem use the conformal flatness approximation in
constructing the metric.  This is problematic because isolated Kerr
black holes are not conformally flat, and forcing them to be so is
tantamount to adding in a compensating gravitational radiation field.
Conformally flat treatments of spinning binary black holes are therefore
contaminated with spurious gravitational radiation, and are probably not
as reliable as treatments of zero spin black holes.

Nevertheless, the calculations of \cite{pfe00}, who generalized the
numerical calculation of \cite{coo94} by allowing for spin, but again
assumed conformal flatness, are revealing. They considered binaries
consisting of holes of equal mass and equal spin magnitude in circular
orbits.  Their study was restricted to binaries in the range $--$0.50 to
$++$0.17.  where the $+$ or $-$ sign denotes whether each hole is co- or
counter-rotating, respectively, and the numerical coefficient indicates
the magnitude of the spin parameter $J/M_{\rm BH}^2$ of each hole.
Assuming that the total mass and angular momentum are conserved during
the plunge from the ISCO, one finds from their numerical data that the
spin $J/M^2$ of the final, merged hole varies from 0.63 for the
$--$0.50 binary to 0.82 for the $++$0.17 binary.

These values can be compared with the strict upper limit, $(J/M^2)_{\rm
max}$, provided by the area theorem, combined with the fact that the
final angular momentum cannot exceed the total angular momentum of the
system at the ISCO:
\begin{equation}
\label{max}
\left({J\over{M^2}}\right)_{\rm max} = {2 \over x_{\rm max}}
{\left ( 1 - {1 \over x^2_{\rm max}} \right )^{1/2}}
\end{equation}
where 
\begin{equation}
\label{irr}
x_{max}^2 = 1 + {{\bar J}^2 \over
4 \left( 1 + \sqrt {1 - \left (J/M_{\rm BH}^2 \right)} \right )^2}
\end{equation}
\citep{pct02}.  The values of $(J/M^2)_{\rm max}$ reach 0.92 for the $--$0.50
binary and 0.97 for the $++$0.17 binary. We expect that the values given
by assuming that mass and angular momentum at the ISCO are conserved
during the plunge provide more realistic estimates. The reason is that
the amount of radiation allowed by the area theorem is far larger than
actually found by numerical computations, where such computations are
available.  For example, the area theorem allows $29 \%$ of the
mass-energy to be radiated in a head-on collision of identical,
nonspinning  black holes falling from rest at large separation, while
the numerical calculations yield $ 0.1 \%$ (\citealt{sma79,ann93}). 

Typically, therefore, the merger of two black holes of comparable mass
will immediately drive the spin parameter of the merged hole to $\gta
0.8$.

\section{Spin-Down by Minor Mergers}\label{minor}

Recently, \cite{hb03} have considered the spin evolution of a black
hole due to mergers with smaller companions.  Here we briefly revisit
the problem.  Our approach is slightly simplified, does not use the
Fokker-Planck formalism, and evaluates the power law for spin decay
exactly in the limit of small $j$.  Along the way, we provide a small
$j$ expansion for the radius and specific energy of the ISCO.

Consider a merger between a large black hole of mass $M$ and a small
black hole of mass $m$, with $q \equiv m/M \ll 1$.  The large black hole
has spin angular momentum $\bS$.  The change in the total spin angular
momentum of the black hole is
\begin{equation}
\Delta (\bS^2) = (\bS + \Delta \bS)^2 - \bS^2 = 2 \bS \cdot \Delta \bS +
	(\Delta \bS)^2.
\end{equation}
We are interested in how the spin evolves due to a large number of
mergers, so we take an ensemble average:
\begin{equation}
\<\Delta (\bS^2)\> = \< 2 \bS \cdot \Delta \bS \> + \< (\Delta \bS)^2 \>.
\end{equation}
The first term on the right is the ``resistive'' or ``dynamical
friction'' term, due to correlations between the black hole spin and its
change in angular momentum due to the accreted object.  The second term
on the right is the ``random walk'' term.  What is perhaps surprising is
that the dynamical friction term does not vanish.

To evaluate $\Delta \bS$ we will assume that the merger occurs through
the slow, gravity-wave driven inspiral of the smaller hole onto the
larger hole.  The larger hole is assumed to be slowly rotating, $j \ll
1$, and the smaller hole is assumed to have isotropically distributed
orbital angular momentum $\bL$.  The inspiral may be regarded as
progressing through a series of nearly circular orbits of fixed
inclination \citep{hug01}.  The change in total angular momentum of the
larger black hole is then approximately $\bL$ evaluated at the ISCO (we
assume that the radiation of energy and angular momentum during the
plunge is negligible).

With this picture in mind, the random walk term is
\begin{equation}\label{randwalk}
\< (\Delta \bS)^2 \> \approx \bL^2 = (m l M)^2,
\end{equation}
where $l \approx 2 \sqrt{3} + \order(j)$ is the specific orbital angular
momentum of a particle on the ISCO around a hole of unit mass.

Now consider the dynamical friction term.  One might naively expect
$\bL$ to be uncorrelated with $\bS$ if $\bL$ is isotropically
distributed, as we have assumed.  But consider a particle with orbital
inclination angle $i$, $\mu = \cos(i)$.  By expanding the fundamental
equations for the ISCO to lowest order in $j$ (see \citealt{hb03}
equations 1 and 2), one can show that
\begin{equation}
{L_z\over{m M}} = 2 \sqrt{3}\, \mu - {2\sqrt{2}\over{3}}\, j \mu^2 +
\order(j^2),
\end{equation}
\begin{equation}
\sE = {2 \sqrt{2}\over{3}} - {1\over{18 \sqrt{3}}}\, j \mu + \order(j^2),
\end{equation}
and
\begin{equation}
{r\over{M}} = 6 -  {4 \sqrt{2}\over{\sqrt{3}}} \, j \mu + \order(j^2).
\end{equation}
Here $L_z$ is the component of orbital specific angular momentum
parallel to the black hole spin axis, $\sE$ is the particle specific
energy, and $r$ is the ISCO orbital radius measured in Boyer-Lindquist
coordinates.  Evidently prograde ($\mu = 1$) orbits have orbital angular
momentum of smaller magnitude than retrograde ($\mu = -1$) orbits; their
ISCO lies ``closer'' to the black hole.  Thus $\bS$ is correlated with
$\Delta \bS$.

Using this result, 
\begin{equation}
2 \bS \cdot \Delta \bS = 2 j M^2 \Delta \bS_z \approx 2 j M^3 m
(2 \sqrt{3} \mu - {2 \sqrt{2}\over{3}} \mu^2 j).
\end{equation}
Averaging over $\mu$, 
\begin{equation}
\< 2 \bS \cdot \Delta \bS \> \approx {1\over{2}} \int_{-1}^{1} d\mu
\left( 2 j M^3 m (2 \sqrt{3} \mu - {2 \sqrt{2}\over{3}} \mu^2 j)\right)
= -{ 4 \sqrt{2} M^3 m j^2\over{9}},
\end{equation}
since the first term vanishes under integration.

We are now in a position to develop an expression for the evolution of
$j$:
\begin{equation}
{M \<\Delta (\bS^2)\>\over{\sE m}} 
\approx {M d(j^2 M^4)\over{ d M }}
= {d (j^2 M^4)\over{ d\ln M }} 
\end{equation}
where $\sE m$ is the change in mass of the large hole in each event,
and we can pass to the continuum limit only if $q \ll 1$.  Using
equation (\ref{randwalk}),
\begin{equation}
{d (j^2 M^4)\over{ d\ln M }} 
= {M\over{\sE m}} \left(
-{4 \sqrt{2}\over{9}} j^2 M^3 m + l^2 m^2 M^2 \right),
\end{equation}
and solving for $d \ln j/d\ln M$,
\begin{equation}
{d \ln j \over{d \ln M}} = - 2 - {2 \sqrt{2}\over{9 \sE}}
	+ {l^2 m\over{2 \sE M j^2}}.
\end{equation}
The first term describes conservation of spin angular momentum (Hughes
\& Blandford's ``doctrine of original spin''),  the second term
dynamical friction, and the third term the random walk.  Substituting
for $\sE$ and $l$,
\begin{equation}
{d \ln j \over{d \ln M}} = - {7\over{3}} + {9 m\over{\sqrt{2} M j^2}}.
\end{equation}
The final term can be ignored whenever $j^2/q \gg 27/(7 \sqrt{2})$; then
$j \sim M^{-7/3}$, in agreement with the \cite{hb03} result $j \sim
M^{-2.4}$.  At late times, when the final term becomes comparable to the
first term, the hole will fluctuate around $j \sim q^{1/2}$.  Thus minor
mergers with smaller objects with isotropically distributed orbital
angular momentum will spin down a hole.

\section{Spin-up By Gas Accretion}\label{accretion}

Once formed, black holes may grow through accretion of the surrounding
plasma.  Accretion onto GBHCs in X-ray binaries is well established. 
In the case of SMBHs,
appreciable growth of black hole seeds by gas accretion is
supported by the consistency between the total energy density in QSO
light and the BH mass density in local galaxies, adopting a reasonable
accretion rest-mass--to--energy conversion efficiency
(\citealt{sol82,yt02,erz02}).  But quasars have been discovered out to
redshift $z \sim 6$, so it follows that the first SMBHs must have formed
by $z_{\rm BH} \gtrsim 6$ or within $t_{\rm BH} \lesssim 10^9$ yrs after
the Big Bang. This timescale provides a tight constraint on SMBH seed
formation scenarios.  For example, it has been argued that if they
indeed grew by accretion, seeds of mass $ \gtrsim 10^5 {\rm ~M_{\odot}}$
must have formed by $ z \sim 9$ to have sufficient time to reach a mass
of $ \sim 10^9 {\rm ~M_{\odot}}$ \citep{gne01}. For a discussion of 
plausible scenarios for forming these black hole seeds, see \cite{shap03}.

Accretion will cause the spin $j = a/M = J/M^2$ of a black hole to
evolve in both magnitude and direction.  In this section we will assume
that the orientation of the spin vector is fixed.  Adopting Kerr-Schild
coordinates $t,r,\theta,\phi$, the angular momentum accretion rate is
\begin{equation}
\dot{J} \equiv \int d\theta d\phi\, \sqrt{-g}\, {T^{r}}_\phi
\end{equation}
where the integral is taken on the horizon, $g$ is the metric
determinant, and ${T^\mu}_\nu$ is the stress-energy tensor of the
accreting material.\footnote{We assume that the mass and angular
momentum of the disk is small compared to that of the black hole.}  We
will consistently use Kerr-Schild coordinates here and below, but notice
that Kerr-Schild $r$ and $\theta$ are identical to the more familiar
Boyer-Lindquist $r$ and $\theta$.  Mass-energy is accreted at a rate
\begin{equation}
\dot{M} = \dot{E} \equiv \int d\theta d\phi\, \sqrt{-g}\, {T^{r}}_t,
\end{equation}
and finally spin evolution is governed by
\begin{equation}
{d j\over{d t}} = {\dot{J}\over{M^2}} - {2 j \dot{E}\over{M}}.
\end{equation}
It is useful to define the dimensionless spinup parameter $s$, 
\begin{equation}
s \equiv {d j\over{d t}} {M\over{\dot{M}_0}},
\end{equation}
where
\begin{equation}
\dot{M}_0 \equiv \int d\theta d\phi \, \sqrt{-g} \, \rho_0 u^r
\end{equation}
is the rest-mass accretion rate, $\rho_0$ is the rest-mass density, and
$u^r$ is the radial component of the plasma four-velocity.  If $s < 0$
the black hole is spinning down.

It was first noted by \cite{bar70} that a black hole accreting 
through a cold disk of constant orientation could achieve maximal
rotation $j = 1$ in finite time.  Bardeen assumed that $\dot{J} = l
\dot{M}_0 M$ and $\dot{E} = \sE \dot{M}_0$ where $l = u_{\phi}$ and $\sE =
-u_t$ are the mean angular momentum and energy per unit rest mass of a
particle on the ISCO, respectively.  This is equivalent to the ``no
torque boundary condition'': no torque is exerted on the disk by the
plasma in the plunging region.

\cite{tho74} noted that a thin disk inevitably radiates, and some of
this radiation will be accreted by the hole.  Preferential accretion of
low angular momentum photons then limits the spin of the hole to a
maximum value $j = 0.998$.  

Exceptions to the \cite{tho74} result have been noted by several authors
in succeeding years.  \cite{ajs78} pointed out that material accreted
from a thick, partially pressure supported disk would have {\it greater}
specific angular momentum than that accreted from a thin disk, and so
one might obtain $j > 0.998$.  \cite{pg98} considered the evolution of a
relativistic hot flow using a viscous model for angular momentum
transport.  They typically found spin equilibrium at $j \simeq 0.7$.
Finally, \cite{tho74}, quoting Bardeen, noted that magnetic fields could
connect material in the disk and the plunging region, and thus sharply
reduce the equilibrium spin.

Magnetic fields in the plunging region have received some attention in
recent years, beginning with the work of \cite{kro99} and \cite{gam99},
based on earlier work by \cite{tak90} and others.  The inflow model of
\cite{gam99} considered an inflowing plasma near the equatorial plane.
It assumed that a well-ordered magnetic field threaded both the plunging
region and the inner edge of the disk, and integrated the resulting one
dimensional steady flow equations.  \cite{li00,li02} considered a model
in which a thin disk is connected to the black hole via high latitude
field lines, rather than through the inflow itself.  \cite{ak00}
considered the implications of the magnetic accretion torques, including
the possible change in surface brightness of the disk.

Recently it has become possible to study the dynamics of nonspherical
black hole accretion numerically in full general relativity
\citep{gam03,dvh03}.  Given an appropriate set of initial and boundary
conditions, this allows one to calculate $dj/dt$ directly for a
nonradiative flow, within the magnetohydrodynamic (MHD) approximation.

Here we consider a sequence of initial conditions in which $j$ alone is
varied and ask for which model is spin equilibrium, $s = 0$, achieved.
We use HARM \citep{gam03} to integrate the equations of ideal, general
relativistic magnetohydrodynamics (GRMHD) in a stationary, Kerr
background spacetime in a variant of Kerr-Schild coordinates.
Axisymmetry is assumed.  Details of the scheme, the coordinate system,
and an extensive series of convergence tests may be found in
\cite{gam03}.

Our initial conditions contain a \cite{fm76} torus with inner radius at
$r = 6 M$ and pressure maximum at $r = 12 M$.  The initial magnetic
field is purely poloidal and field lines follow isodensity contours.  It
is derived from a vector potential $A_\phi \propto {\rm
MAX}(\rho_0/\rho_{0,min} - 0.2, 0)$.  The field is constrained to have a
minimum ratio of gas to magnetic pressure of $100$.  Our equation of
state is $p = (\gamma - 1) u$, where $p$ is the gas pressure, $u$ is the
internal energy, and $\gamma = 4/3$.  

We set the inner boundary of the computational domain at $r_{in} = 0.98
r_h$ where $r_h = (1 + \sqrt{1 - j^2})$ is the event horizon radius.
Because the inner boundary is inside the event horizon it is causally
isolated from the rest of the flow.  The outer boundary of the
computational domain is located at $r_{out} = 40 M$.  This is distant
enough that the influence of the outer boundary on the inner accretion
flow ($r \sim M$) is negligible.  Outflow boundary conditions (zero
order extrapolation of primitive variables with a switch forbidding
inflow) are used at the outer boundary.  Tests indicate that all results
are independent of $r_{in}$ and $r_{out}$, unless $r_{in}$ is outside
the horizon.  Our numerical resolution is $256 \times 256$ in zones
equally spaced in the coordinates $x_1 = \ln r$ and $x_2$ such that
$\theta = \pi x_2 + (1/2) (1 - H) \sin(2\pi x_2)$, where $H$ is a
parameter that gradually concentrates zones toward to the equator as $H$
is decreased from $1$ to $0$.  Here we use $H = 0.3$.  We find that
varying the resolution by a factor of $2$ in either direction leaves our
results unchanged.

Figure 1 illustrates the outcome of a typical evolution of this
configuration around a black hole with $j = 0.75$.  It shows the
logarithm of the density field in the $R = r \cos(\theta), Z = r
\sin(\theta)$ plane.  The disk has become turbulent due to the
magnetorotational instability \citep{bh91}, and angular momentum
transport by MHD turbulence leads to gradual inflow along the equator.
The overall structure of the flow is similar to that observed by
\cite{dvhk}, with an evacuated funnel near the poles, outflow at
intermediate latitudes, a nearly-Keplerian equatorial torus, and a
plunging region between the torus and the event horizon.

Figure 2 shows the evolution of $\dot{M}_0, \dot{E}/\dot{M}_0$, and
$\dot{J}/(\dot{M}_0 M)$ for a model with $j = 0.9375$.  The cyan lines
show the values expected for the classical thin disk in which the
specific energy and angular momentum of accreted material is equal to
that of a particle on the ISCO, as in \cite{bar70}.  While the specific
energy of accreted material is accurately predicted by the thin disk
model, the specific angular momentum is substantially lower.  This is a
result of ordered magnetic fields in the plunging region which transport
angular momentum outward into the bulk of the disk.  

A similar suppression of accreted angular momentum has been observed in
fully relativistic MHD simulations of accretion onto black holes by
\cite{dvh} and \cite{dvhk}.  In particular, Table 2 of the latter can be
used to estimate $s = 1.64$ for their model with $j = 0.5$.  This is in
very close agreement with $s = 1.66$ for our model.  Given that
different initial conditions and completely different numerical methods
were used for these two calculations, and that our calculation is two
dimensional while de Villiers et al.'s calculation is three dimensional,
the agreement is remarkable.  We also note that the $j = 0.998$
calculation by \cite{dvhk} shows $s < 0$.

Figure 3 shows the dimensionless spin evolution factor $s =
(dj/dt)(M/\dot{M}_0)$ for four separate models with $j = 0.5, 0.75,
0.88, 0.97$.  At $j = 0.97$ the hole is spinning down ($s < 0$).  Figure
4 shows $s(j)$, including models that were excluded from Figure 3 for
clarity.  This sequence of accretion models reaches equilibrium ($s =
0$) for $j \simeq 0.93$.  This suggests that magnetic interactions can
lead to spin equilibration at lower $j$ than the canonical $j = 0.998$
of \cite{tho74}.

We are not suggesting that spin equilibrium is {\it always} reached at
$j \simeq 0.93$.  Models with different initial conditions may produce
different results.  For example, the models we have considered here have
a ratio of scale height to local radius $H/r \sim 0.2$--$0.3$ at
pressure maximum.  Thinner disks are likely to produce different
results; indeed, in a later publication we will present results from the
self-consistent evolution of a thin disk that closely match the
predictions of a classical thin disk model.  But thinner disks imply
lower accretion rates, so thin disk accretion will have lower weight in
determining the black hole spin than thick disk accretion over a
comparable timescale.  To sum up, what we have shown is that there exist
self-consistent GRMHD models with $j \gta 0.93$ that are unambiguously
spinning down the black hole: accretion need not necessarily lead to
near-maximal rotation.

\section{Conclusions}\label{finale}

In this paper, we have considered astrophysical processes that influence
the spin evolution of black holes.  Fully relativistic collapse
calculations suggest that the initial spin of a newborn black hole $j
\lesssim 0.75$--$0.9$, where the upper limit applies to the collapse of
maximally and uniformly rotating massive stars.  The outcome of
subsequent mergers with black holes of comparable mass is not yet fully
understood, but current best estimates suggest a final spin $j \sim
0.8$--$0.9$ (see Table 1).  Mergers with black holes of much smaller
mass can be treated in the test-particle approximation and, following
\cite{hb03}, we have presented an argument showing that such mergers
tend to spin down the black hole to $j \ll 1$, provided that the small
black holes have isotropically distributed orbital angular momentum.

We have also presented results from fully relativistic
magnetohydrodynamic models of accretion onto a rotating hole.   These
show that, at least for the particular series of thick disk models we
consider, spin equilibrium is reached at $j \approx 0.93$.  This
demonstrates that accretion need not lead to near-maximal rotation.  Our
models have a ratio of scale height to local radius $H/r \sim
0.2$--$0.3$, and thus correspond to near-Eddington accretion rates.
Accretion at lower rates (through thinner disks) may be capable of
producing higher spin, provided that the orbital angular momentum of the
accreting material remains aligned with the black hole spin.  If the
orbital plane of the accreting material varies, as seems likely (see the
discussion of \citealt{np98}), even thin disk accretion may be unable to
produce $j \approx 1$.

All these results suggest that near-maximal rotation of black holes is
neither necessary nor likely.  Black hole spins $j \sim 0.7$ -- $0.95$
are produced in a variety of scenarios.  This corresponds to thin disk
radiative efficiencies of $10\%$--$19\%$, which is broadly consistent
with the radiative efficiencies required by Soltan-type arguments
\citep{yt02,erz02}.  Such modest spins are also not in conflict with the
idea that radio galaxies are powered by black hole spindown.  The
\cite{bz77} luminosity of a black hole scales as $j^2 (B^r)^2$ where
$B^r$ is the mean radial magnetic field on the event horizon.  Unless
$B^r$ is a sharply increasing function of $j$, the Blandford-Znajek
luminosity of a black hole with $j \simeq 0.9$ is not very different
from that of a nearly maximally rotating black hole.  Only if black
holes were built up mainly through thin disk (sub-Eddington) accretion
of material in a fixed orbital plane would near-maximal rotation be the
norm.

\acknowledgments

This work was supported in part by NSF ITR Grant PHY 02-05155, NSF
PECASE Grant AST 00-93091, NSF Grant PHY-0090310, NASA Grant NAG5-8418
and NASA Grant NAG5-10881 to the University of Illinois at
Urbana-Champaign.  JCM was supported by a NASA GSRP fellowship.  CFG is
pleased to acknowledge a National Center for Supercomputing Applications
(NCSA) faculty fellowship.  Some of the computations described here were
performed at NCSA.

\clearpage

\begin{figure}
\epsscale{0.5}
\plotone{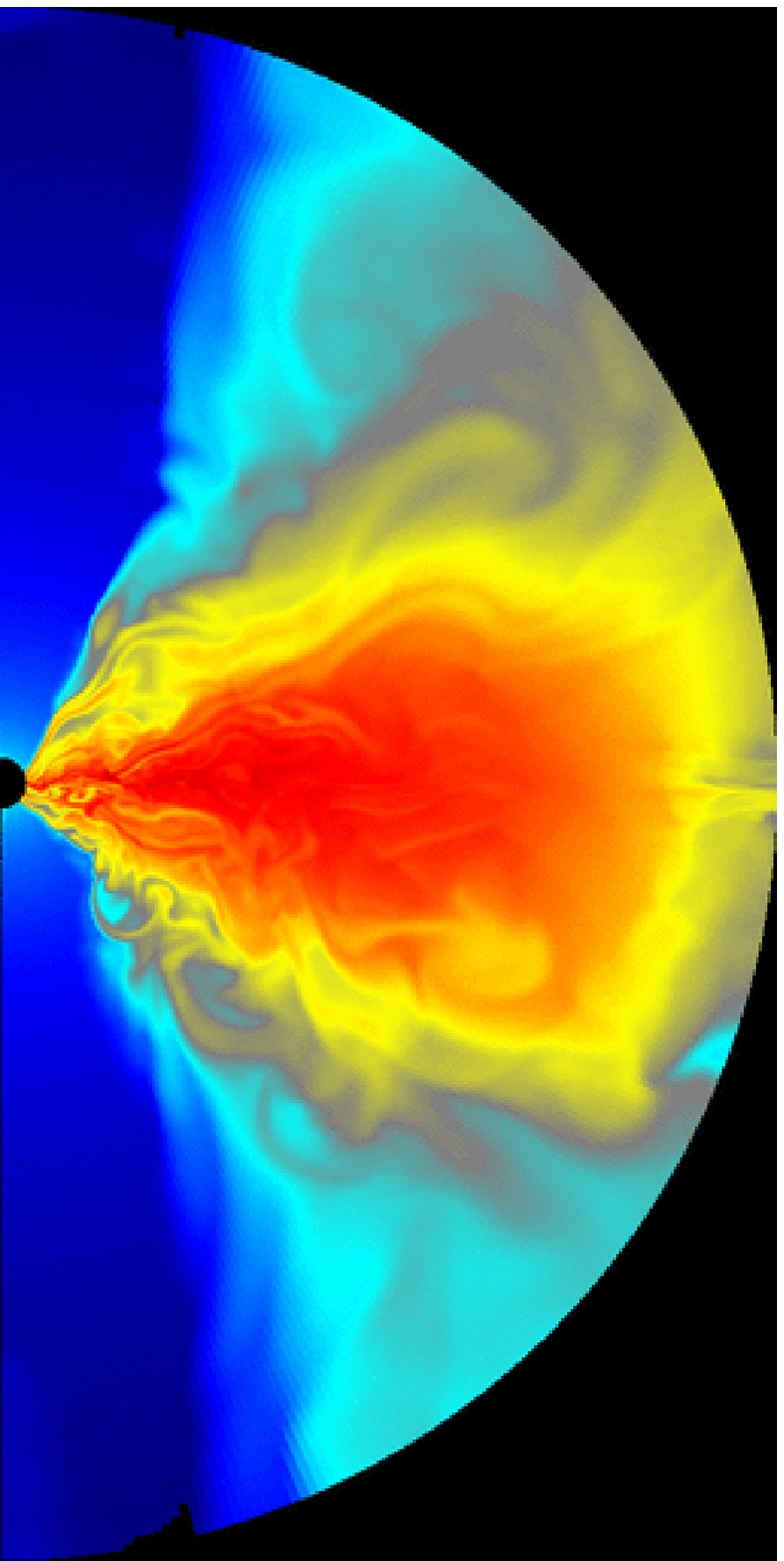}
\caption{
A snapshot from an evolution of a weakly magnetized Fishbone-Moncrief
torus around a $j = 0.75$ black hole.  Color corresponds to
$\log(\rho_0)$.  Red is high rest-mass density, and black is low.
}
\end{figure}

\begin{figure}
\plotone{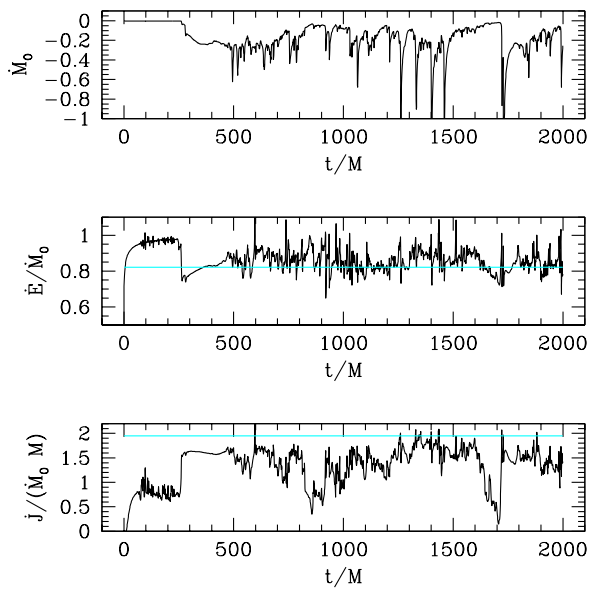}
\caption{
Evolution of the mass, energy, and angular momentum accretion rate for a
weakly magnetized Fishbone-Moncrief tori around a black hole with $j =
0.9375$.
}
\end{figure}

\begin{figure}
\plotone{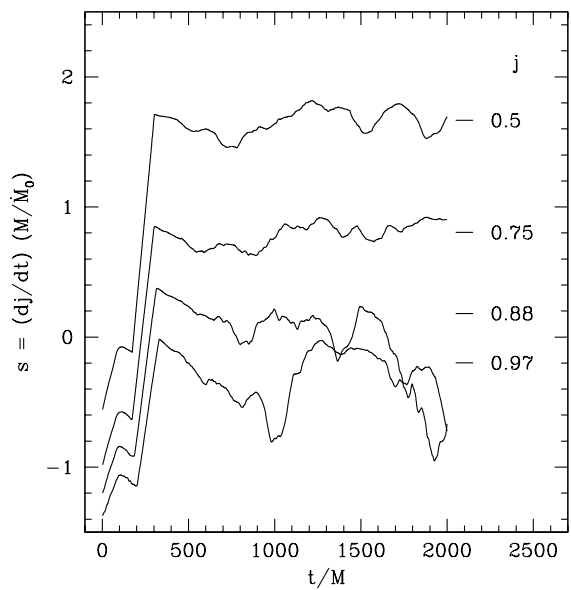}
\caption{
Evolution of $s = (dj/dt)(M/\dot{M}_0)$ for a series of four
Fishbone-Moncrief tori.
}
\end{figure}

\begin{figure}
\plotone{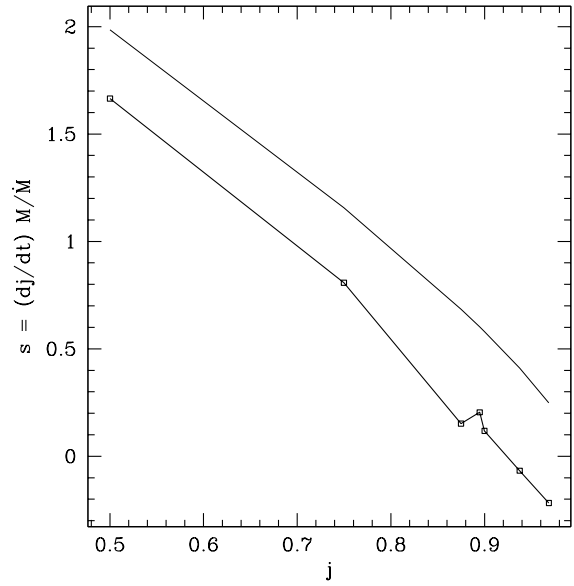}
\caption{
Time-averaged value of $s$ for a sequence of Fishbone-Moncrief tori.
The squares indicate data points from simulations, while the thin line
indicates values expected for a thin disk.
}
\end{figure}

\clearpage

\begin{table}
\begin{center}
\caption{Representative values for nonspinning binary black holes}
\begin{tabular}{llllll}
\tableline
\tableline
Reference                   & $\bar E_b$\tablenotemark{a} & $ \bar
J$\tablenotemark{b}     &$\bar \Omega$\tablenotemark{c}
 &$ J/M^2$\tablenotemark{d}      &$(J/M^2)_{\rm max}$\tablenotemark{e}       \\
\hline
Schwarzschild               & -0.0572       & 3.464         & 0.068
 & 0.8913       & 0.9897\\
Cook(1994)                  & -0.09030      & 2.976         & 0.172
 & 0.7788       & 0.9578\\
Baumgarte (2000)            & -0.092        & 2.95          & 0.18
 & 0.773        & 0.955\\
Grandclement et al. (2002)  & -0.068        & 3.36          & 0.103
 & 0.869        & 0.985\\
Damour et al. (2000)        & -0.0668       & 3.27          & 0.0883
 & 0.846        & 0.980\\
\tableline
\end{tabular}
\tablenotetext{a}{Binding energy per unit reduced mass at the ISCO.}
\tablenotetext{b}{Angular momentum per unit reduced mass at the ISCO.}
\tablenotetext{c}{Orbital angular velocity at the ISCO.}
\tablenotetext{d}{Estimated spin parameter of final black hole.}
\tablenotetext{e}{Maximum  spin parameter of final black hole (see
text).
}
\end{center}
\end{table}

\clearpage


\clearpage

\begin{thebibliography}{}

\bibitem[Abramowicz, Jaroszinski, \& Sikora(1978)]{ajs78} Abramowicz,
        M., Jaroszinski, M., \& Sikora, M. 1978, AA, 63, 221

\bibitem[Agol \& Krolik(2000)]{ak00} Agol, E.~\& Krolik, J.~H.\ 2000,
\apj, 528, 161 

\bibitem[Anninos(1993)]{ann93} Anninos, P., Hobill, D., Seidel, E., 
Smarr, L., Suen, W.-M. 1993, Phys. Rev. Lett, 71, 2851

\bibitem[Balbus \& Hawley(1991)]{bh91} Balbus, S.A., \& Hawley, J.F.
        1991, ApJ, 376, 214

\bibitem[Bahcall et al.(1990)]{bahc90} Bahcall, S., Lynn, B.W., \&
Selipsky, S.B. 1990, ApJ, 362, 251

\bibitem[Bardeen(1970)]{bar70} Bardeen, J.M. 1970, Nature, 226, 64

\bibitem[Baumgarte(2000)]{bau00} Baumgarte, T.W. 2000, Phys. Rev. D.,
62, 084020

\bibitem[Baumgarte \& Shapiro(1999)]{baum99} Baumgarte, T.W., \& Shapiro,
S.L. 2003, ApJ, 526, 941

\bibitem[Baumgarte \& Shapiro(2003)]{bs03} Baumgarte, T.W., \& Shapiro,
S.L. 2003, Phys. Rep. 376, 41

\bibitem[Blandford \& Znajek(1977)]{bz77} Blandford, R.D., \& Znajek, R.
        1977, MNRAS, 179, 433

\bibitem[Christodoulou(1970)]{chr70} Christodoulou, D. 1970 Phys. Rev. Lett., 25, 1596

\bibitem[Cook(1994)]{coo94} Cook, G.B. 1994, Phys. Rev. D., 50, 5025

\bibitem[Cook(2003)]{cook03} Cook, G. B., talk presented at KITP Conference
on Gravitational Interaction of Compact Objects, May 13 2003, 
($http://online.itp.ucsb.edu/online/gravity\_c03/cook/$).

\bibitem[Damour et al.(2000)]{djs00} Damour, T., Jaranowski, P.,
\& Sch\"afer, G. 2000, Phys. Rev. D. 62, 084011

\bibitem[DeDeo \& Psaltis(2003)]{dp03} DeDeo, S., \& Psaltis, D. 2003,
astro-ph/0302095

\bibitem[Dreyer et al.(2003)]{drey03} Dreyer, O., Kelly, B., Krishnan,
B., Finn, L.S., Garrison, D., Lopez-Aleman, R. 2003, gr-qc/0309007

\bibitem[Gezari et al.(2002)]{gez02} Gezari, S., Ghez, A.~M., Becklin,
E.~E., Larkin, J., McLean, I.~S., \& Morris, M.\ 2002, \apj, 576, 790

\bibitem[Gierli{\' n}ski, Macio{\l}ek-Nied{\' z}wiecki, \&
Ebisawa(2001)]{gme01} Gierli{\' n}ski, M., Macio{\l}ek-Nied{\' z}wiecki,
A., \& Ebisawa, K.\ 2001, \mnras, 325, 1253

\bibitem[Grandecl\'ement et al.(2002)]{ggb02} Grand\-cl\'{e}ment, P.,
Gourgoulhon, E., \& Bonazzola, S.,  2002, Phys. Rev. D., 65, 044021

\bibitem[Heger et al.(2002)]{heg02} Heger, A., Woosley, S., Baraffe, I.,
\& Abel, T.\ 2002, Lighthouses of the Universe: The Most Luminous
Celestial Objects and Their Use for Cosmology Proceedings of the
MPA/ESO/, p.~369, 369 

\bibitem[Elvis, Risaliti, \& Zamorani(2002)]{erz02} Elvis, M.,
        Risaliti, G., \& Zamorani, G.\ 2002, \apjl, 565, L75

\bibitem[Evans \& Hawley(1988)]{eh88} Evans, C.R., \& Hawley, J.F. 1988,
        \apj, 332, 659

\bibitem[Fabian et al.(2002)]{fab02} Fabian, A.~C.~et al.\ 
2002, \mnras, 335, L1

\bibitem[Fishbone \& Moncrief(1976)]{fm76} Fishbone, L.G., \& Moncrief,
        V. 1976, ApJ, 207, 962

\bibitem[Font(2000)]{font00} Font, J. A., 2000, Liv. Rev. in Rel., 3,
        2000-2font

\bibitem[Flanagan \& Hughes(1998)]{flan98} Flanagan, E. E., \& Hughes, S. A.,
1998, \prd, 57, 4535

\bibitem[Gammie(1999)]{gam99} Gammie, C.F. 1999, ApJL, 522, L57

\bibitem[Gammie, McKinney, \& T{\' o}th(2003)]{gam03} Gammie, C.~F.,
McKinney, J.~C., \& T{\' o}th, G.\ 2003, \apj, 589, 444


\bibitem[Gnedin(2001)]{gne01} Gnedin, O. Y. 2001, Class. \& Quan. Grav., 18, 3983

\bibitem[Harten et al.(1983)]{hll83} Harten, A., Lax, P.D., \& van Leer,
        B. 1983, SIAM Rev. 25, 35

\bibitem[Hawley, Smarr, \& Wilson(1984)]{hsw} Hawley, J.F., Smarr, L.L.,
        \& Wilson, J.R. 1984, ApJS, 55, 211

\bibitem[Hughes(2001)]{hug01} Hughes, S.~A.\ 2001, \prd, 64, 64004 

\bibitem[Hughes \& Blandford(2003)]{hb03} Hughes, S.~A.~\& Blandford,
R.~D.\ 2003, \apjl, 585, L101 

\bibitem[Khanna et al.(1999)]{kh99} Khanna, G., Baker, J., Gleiser, R.,
Laguna, P., Nicasio, C., Nollert, H-P., Price, R., \& Pullin, J.
1999, \prl, 83, 3581

\bibitem[Krolik(1999)]{kro99} Krolik, J.~H.\ 1999, \apjl,
515, L7

\bibitem[Laor(1991)]{lao91} Laor, A.\ 1991, \apj, 376, 90 

\bibitem[Li(2000)]{li00} Li, L.\ 2000, \apjl, 533, L115 

\bibitem[Li(2002)]{li02} Li, L.\ 2002, \apj, 567, 463 

\bibitem[MacFadyen \& Woosley(1999)]{mw99} MacFadyen, A. I. \& Woosley, S. E.\
1999, \apj, 524, 262

\bibitem[Magorrian et al.(1998)]{mag98} Magorrian, J.~et al.\ 1998, \aj,
115, 2285 

\bibitem[Maoz(1998)]{mao98} Maoz, E.\ 1998, \apjl, 494, L181 

\bibitem[McClintock \& Remillard(2003)]{mr03} McClintock, J.E., \&
Remillard, R.A. 2003, astro-ph/0306213

\bibitem[Merritt \& Eckers(2002)]{me02} Merritt, D. \& Eckers, R. D.
2002, Science, 297, 1310

\bibitem[Miller et al.(2002)]{mil02a} Miller, J.~M.~et al.\ 
2002, \apj, 578, 348 

\bibitem[Miller et al.(2002)]{mil02b} Miller, J.~M.~et al.\ 
2002, \apjl, 570, L69 

\bibitem[Miller et al.(2001)]{mil01} Miller, J.~M.~et al.\ 
2001, \apj, 563, 928 

\bibitem[Miyoshi et al.(1995)]{miy95} Miyoshi, M., Moran,
J., Herrnstein, J., Greenhill, L., Nakai, N., Diamond, P., \& Inoue, M.\
1995, \nat, 373, 127 

\bibitem[\'{O}Murchadha and York (1974)]{my74} \'{O}Murchadha, N.
\& York Jr., J. W., 1974 Phys. Rev. D. 10, 2345 

\bibitem[Natarajan \& Pringle(1998)]{np98} Natarajan, P.~\& Pringle,
J.~E.\ 1998, \apjl, 506, L97 

\bibitem[Pfeiffer et al.(2000)]{pfe00} Pfeiffer, H. P., Cook, G.B.,
\& Teukolsky, S. A. 2002, Phys. Rev. D., 62, 104018 

\bibitem[Pfeiffer, Cook, \& Teukolsky(2002)]{pct02} Pfeiffer, 
H.~P., Cook, G.~B., \& Teukolsky, S.~A.\ 2002, \prd, 66, 24047 

\bibitem[Popham \& Gammie(1998)]{pg98} Popham, R.~\& Gammie, C.~F.\
1998, \apj, 504, 419 

\bibitem[Reynolds \& Nowak(2002)]{rn02} Reynolds, C.S., \& Nowak, M.A.
2002, Phys. Rep. 377, 389

\bibitem[Sch{\" o}del et al.(2002)]{sch02} Sch{\" o}del, R.~et al.\
2002, \nat, 419, 694

\bibitem[Shahbaz et al.(1999)]{sha99} Shahbaz, T., van der Hooft, F.,
Casares, J., Charles, P.~A., \& van Paradijs, J.\ 1999, \mnras, 306, 89 

\bibitem[Shapiro (2003)]{shap03} Shapiro, S. L. 2003, in Carnegie
Observatories Astrophysics Series, Vol 1: Coevolution of Black Holes and
Galaxies, ed. L. C. Ho (Cambridge: Cambridge Univ. Press), in press
astro-ph/ 0304202.

\bibitem[Shapiro \& Teukolsky(1983)]{st83} Shapiro, S.L., \& Teukolsky,
S.A. 1983, Black Holes, White Dwarfs, and Neutron Stars: The Physics of
Compact Objects (New York: Wiley)

\bibitem[Shapiro \& Shibata(2002)]{shap02} Shapiro, S.~L.~\& Shibata, M.\
2002, \apj, 577, 904 

\bibitem[Shibata \& Shapiro(2002)]{shib02} Shibata, M.~\& Shapiro, S.~L.\
2002, \apjl, 572, L39 

\bibitem[Shibata \& Uryu(2002)]{su02} Shibata, M. \& Uryu, K. 
2002, Prog. Theor. Phys., 107, 265 

\bibitem[Smarr(1979)]{sma79} Smarr, L. 1979, in Sources of 
Gravitational Radiation, ed. L. Smarr (Cambridge University Press, Cambridge)

\bibitem[Soltan(1982)]{sol82} Soltan, A.\ 1982, \mnras, 200, 115

\bibitem[Strohmayer(2001)]{str01} Strohmayer, T.~E.\ 2001, 
\apjl, 552, L49 

\bibitem[Takahashi et al.(1990)]{tak90} Takahashi, M. , Nitta, S. ,
        Tatematsu, Y.  \& Tomimatsu, A.  1990, \apj, 363, 206

\bibitem[Tanaka et al.(1995)]{tan95} Tanaka, Y.~et al.\
1995, \nat, 375, 659 

\bibitem[Thorne(1995)]{tho95} Thorne, K.~S.\ 1995, Seventeeth Texas
Symposium on Relativistic Astrophysics and Cosmology, 759, 127 

\bibitem[Thorne(1974)]{tho74} Thorne, K.~S.\ 1974, \apj, 191, 507 

\bibitem[De Villiers \& Hawley(2003)]{dvh03} De Villiers, J.~\& Hawley,
J.~F.\ 2003, \apj, 589, 458 

\bibitem[de Villiers \& Hawley(2003)]{dvh} de Villiers, J.-P., \&
Hawley, J.F., 2003, astro-ph/0303241

\bibitem[de Villiers et al.(2003)]{dvhk} de Villiers, J.-P., 
Hawley, J.F., \& Krolik, J.H. 2003, astro-ph/0307260

\bibitem[Wald(1984)]{wal84} Wald, R.~M.\ 1984, Chicago:
        University of Chicago Press, 1984, 313

\bibitem[Yu \& Tremaine(2002)]{yt02} Yu, Q.~\& Tremaine, S.\ 2002,
        \mnras, 335, 965

\bibitem[Zhang, Cui, \& Chen(1997)]{zcc97} Zhang, S.~N.,
        Cui, W., \& Chen, W.\ 1997, \apjl, 482, L155


\end{thebibliography}
\end{document}